\documentclass{aa}
\usepackage[varg]{txfonts}
\usepackage{graphicx}
\usepackage{epstopdf}
\usepackage{bm}

\usepackage{natbib}
\bibpunct{(}{)}{;}{a}{}{,} 
\usepackage{threeparttable}
\usepackage{multirow}
\usepackage{xcolor}
\makeatletter
\newcommand*{\rom}[1]{\expandafter\@slowromancap\romannumeral #1@}
\makeatother
\begin{document}

\title{Infrared Spectra of Solid-State Ethanolamine: Laboratory Data in Support of JWST Observations}
\titlerunning{Infrared spectra of solid state Ethanolamine}
\authorrunning{Suhasaria et al.}

   \author{T. Suhasaria \inst{1}, S. M. Wee \inst{1}, R. Basalg{\`e}te \inst{2}, S. A. Krasnokutski
          \inst{2}, C. J{\"a}ger\inst{2},
          G. Perotti\inst{1}
          \and
          Th. Henning\inst{1}
          }

   \institute{Max Planck Institute f\"ur Astronomie, K\"onigstuhl 17, 69117, Heidelberg, Germany\\
   \email{suhasaria@mpia.de}
         \and
             Laboratory Astrophysics Group of the Max Planck Institute for Astronomy at the Friedrich Schiller University Jena, Institute of Solid State Physics, Helmholtzweg 3, 07743, Jena, Germany\\
             }

\abstract
{Ethanolamine (NH$_2$CH$_2$CH$_2$OH, EA) has been identified in the gas phase of the interstellar medium within molecular clouds. Although EA hasn't been directly observed in the molecular ice phase, a solid-state formation mechanism has been proposed. However, the current literature lacks an estimation of the infrared band strengths of EA ices, which are crucial data for quantifying potential astronomical observations and laboratory findings related to their formation or destruction via energetic processing.}
{We conducted an experimental investigation of solid EA ice at low temperatures to ascertain its infrared band strengths, phase transition temperature, and multilayer binding energy. Since the refractive index and the density of EA ice are unknown, the commonly used laser interferometry method was not applied. Infrared band strengths were determined using three distinct methods. Besides the evaluation of band strengths of EA, we also tested the advantages and disadvantages of different approaches used for this purpose. The obtained lab spectrum of EA was compared with the publicly available MIRI MRS James Webb Space Telescope observations toward a low mass protostar.}
{We used a combination of Fourier-transform transmission infrared spectroscopy and quadrupole mass spectrometry.}
{The phase transition temperature for EA ice falls within the range of 175 to 185 K. Among the discussed methods, the simple pressure gauge method provides a reasonable estimate of band strength. We derive a band strength value of about $1\times10^{-17}$ cm molecule$^{-1}$ for the NH$_2$ bending mode in the EA molecules. Additionally, temperature-programmed desorption analysis yielded a multilayer desorption energy of 0.61$\pm$0.01 eV. By comparing the laboratory data documented in this study with the JWST spectrum of the low mass protostar IRAS 2A, an upper-limit for the EA ice abundances was derived.}
{This study addresses the lack of quantitative infrared measurements of EA at low temperatures, crucial for understanding its astronomical and laboratory presence and formation routes. Our approach presents a simple, yet effective method for determining the infrared band strengths of molecules with a reasonable level of accuracy.}

\keywords{astrochemistry -- methods: laboratory: solid state - techniques: spectroscopic - ISM: molecules}
\maketitle
\section{Introduction}

Aminoalcohols are an interesting class of organic compounds that contain both amino (NH$_2$) and hydroxy (OH) functional groups. These molecules, due to their biological relevance, can play an important role in the context of the origin of life. Ethanolamine (NH$_2$CH$_2$CH$_2$OH, hereafter denoted as EA), also known as $\beta$-aminoethanol or glycinol, is the simplest head group of phospholipids and is also considered a precursor of the simplest amino acid, glycine. Additionally, $\alpha$, $\beta$-amino alcohol moieties can be found in various other biomolecules, including aminosugars, sphingolipids, and glycoproteins, playing crucial roles in the functions of biosystems \citep{sladkova2014}.

Recently, EA was identified in the interstellar medium in the molecular cloud G+0.693-0.027, situated within the Sagittarius B2 complex in the Galactic Center, through rotational transition observations as reported by \citet{rivilla2021}. Although EA was observed exclusively in the gas phase and not within interstellar ices, it has been proposed that the gas-phase detection may be attributed to the erosion of icy mantles caused by the low-velocity shocks typically found in massive molecular clouds. EA has also been found in the Almahata Sitta meteorite \citep{glavin2010}.

One of the potential pathways for the formation of EA in space is through the hydrogenation of the aminoketene molecule \citep{rivilla2021}. In a recent study, aminoketene was shown to be formed efficiently at 10 K without the input of energy by the reaction of C atoms with CO and NH$_3$, which could potentially result in high abundances of aminoketene and EA in the solid state \citep{Krasnokutski2021,Krasnokutski2022}. Aminoketene may also polymerize, resulting in the formation of peptide chains that could potentially have a catalytic function, aiding the formation of complex organic or even biological molecules \citep{Krasnokutski2024, Krasnokutski2022}. This pathway begins with the reaction between C and NH$_3$ leading to the formation of HCNH$_2$ and H$_2$CNH products. This differs from a recently published work, which proposes that the reaction of C atoms with an NH$_3$ molecule in ice forms a weakly bound complex, namely CNH$_3$ \citep{Molpeceres24}.

An alternative route to EA is the UV irradiation of molecular ice as demonstrated by laboratory experiments on the photolysis of H$_2$O:CH$_3$OH:NH$_3$:HCN ices, with a 20:2:1:1 mixture \citep{bernstein2002}. In these experiments, EA is produced along with other prebiotic species like the amino acids glycine, alanine, and serine. However, the amount of EA formed in these experiments was very small and it was detected after a chemical treatment. Therefore, the presence of EA, with its multiple functional groups, in molecular ice could lead to the formation of more complex species when exposed to energetic processing, some of which might even have astrobiological significance.

Infrared spectroscopy stands out as the primary method for identifying molecules in extraterrestrial solids, particularly in ice on interstellar and solar system objects. The identification process involves comparing astronomical data to reference measurements from terrestrial laboratories and, in some cases, allows for both identification and quantification of molecular abundances based on IR detections. This method relies on assigning specific IR bands to molecules, integrating them, and comparing the results to intrinsic IR band strengths determined in laboratories.

To our knowledge, there are no quantitative IR measurements of EA ices. This study concentrates on conducting IR spectroscopy on amorphous EA ice at low temperatures and assigning the IR bands based on theoretical calculations. The thermal variation of EA infrared features is also shown. The phase transition temperatures have been identified and the IR spectra of crystalline EA have been highlighted. In addition, experimental band strengths of amorphous EA ice have been determined.

\section{Experimental methods}

Experiments were conducted in the Laboratory Astrophysics group at Jena in the INterStellar Ice Dust Experiment (INSIDE) chamber. The setup has been extensively detailed in prior work \citep{potapov2019}. Here, we provide only a brief overview of the setup which is relevant to this study. The measurements were done in an ultrahigh vacuum (UHV) chamber, maintaining a base pressure better than 8$\times$10$^{-10}$ mbar, which is equipped with a closed-cycle helium cryostat that can cool the sample down to 10 K. The temperature is measured by a diode which is fixed on the sample holder very close to the sample. In our experiments, a potassium bromide (KBr) window was affixed as a substrate on the cold finger of the cryostat. 

Liquid NH$_2$CH$_2$CH$_2$OH (Sigma Aldrich, $\geq$99.0\%) and CH$_3$OH (Sigma Aldrich, $\geq$99.0\%) were purified through multiple freeze-pump-thaw cycles and deposited onto the substrate as pure vapor through a gas inlet system attached to the UHV chamber in different experiments. The gas inlet system has two separate lines. The first line allows mixing of upto five different gases. NH$_2$CH$_2$CH$_2$OH and CH$_3$OH were deposited individually from the first line. The second gas line is kept for corrosive gases. Ammonia (NH$_3$, Linde, 5.5 grade purity) gas was also deposited as a pure component through the second gas line of the same inlet system without any further purification. Gases are deposited onto the substrate through two 0.5 mm diameter capillary tubes. The chamber is equipped with a Fourier transform infrared (FTIR) spectrometer (Bruker Vertex 80v) fitted with a Mercury Cadmium Telluride (MCT) detector for in situ measurement of the ice phase. Infrared spectra were collected in the temperature interval between 10 and 300 K, before and during the warming-up, in transmittance mode in the spectral range from 6000 to 400 cm$^{-1}$ with a resolution of 1 cm$^{-1}$. For the irradiation experiment at 10 K, 256 scans were taken while during the warming-up experiment, only 32 scans were taken each time.

During the low temperature and warming up experiments, a quadrupole mass spectrometer (QMS; HXT300M, Hositrad) attached to the UHV chamber monitors the gas phase inside the chamber to detect molecules desorbing from the sample surface. Temperature programmed desorption (TPD) experiments were performed by linear ramping of the sample temperature with a rate of 10 K minute$^{-1}$ in a temperature range between 10 and 300 K. The error of the temperature measurements was determined to be $\pm$ 1 K.

The molecular geometry and anharmonic vibrational spectrum of one of the conformers of EA were obtained using the second-order Møller–Plesset (MP2) perturbation theory and Aug-cc-pVTZ basis set. Quantum chemical calculations were performed using the Gaussian09 software package \citep{g16}. 

\begin{figure}
\begin{center}
\resizebox{\hsize}{!}{\includegraphics{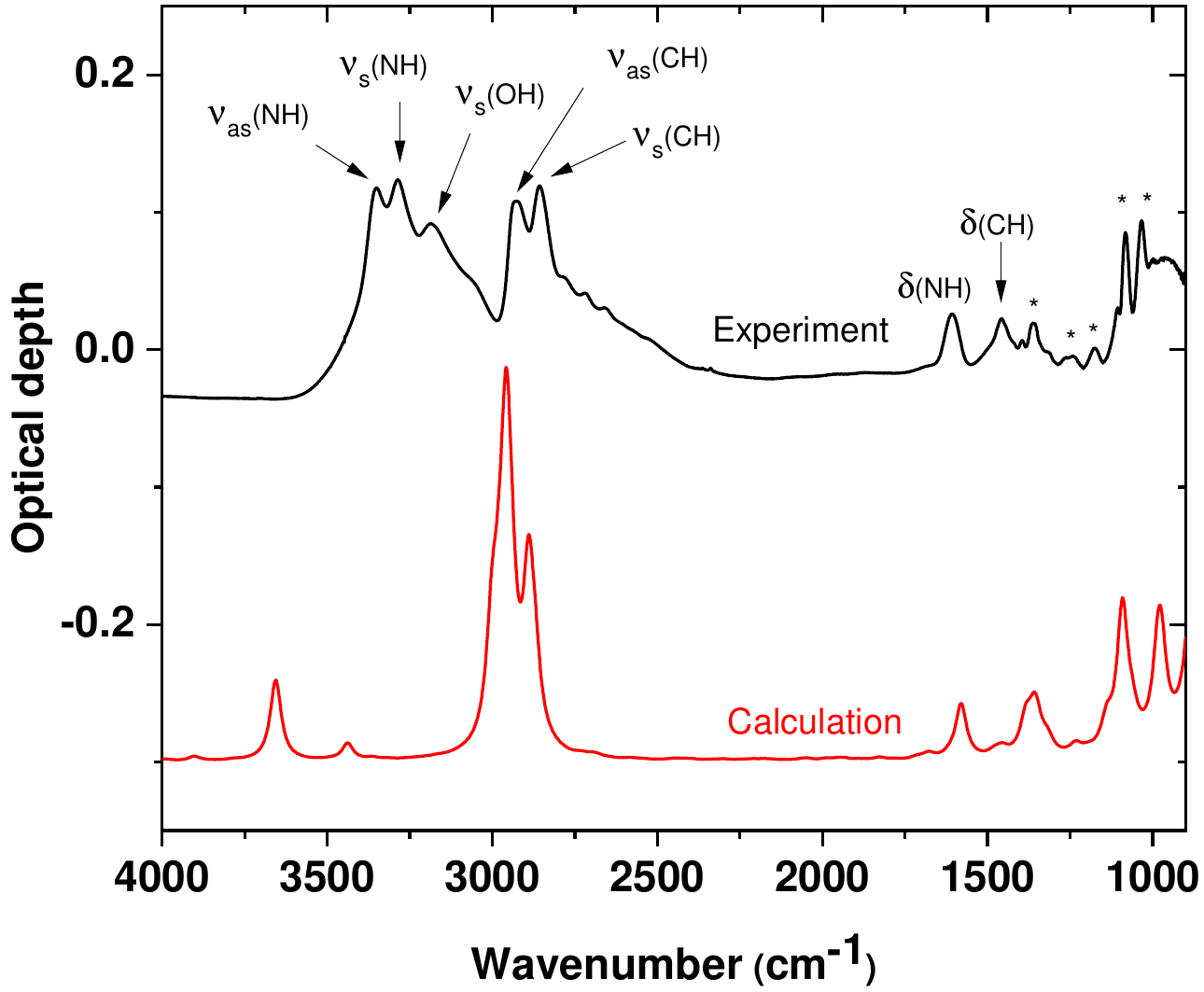}}
\caption{Experimental and theoretical mid-IR spectrum of EA. The upper one shows about 200 monolayers (ML) EA condensed on KBr at 10 K. The infrared bands marked with asterisks are due to a mixture of different vibrational modes. For the latter, calculations were performed at MP2/Aug-cc-pVTZ level.}
   \label{fig1}
\end{center}
\end{figure}

\begin{figure*}
\begin{center}
\resizebox{\hsize}{!}{\includegraphics{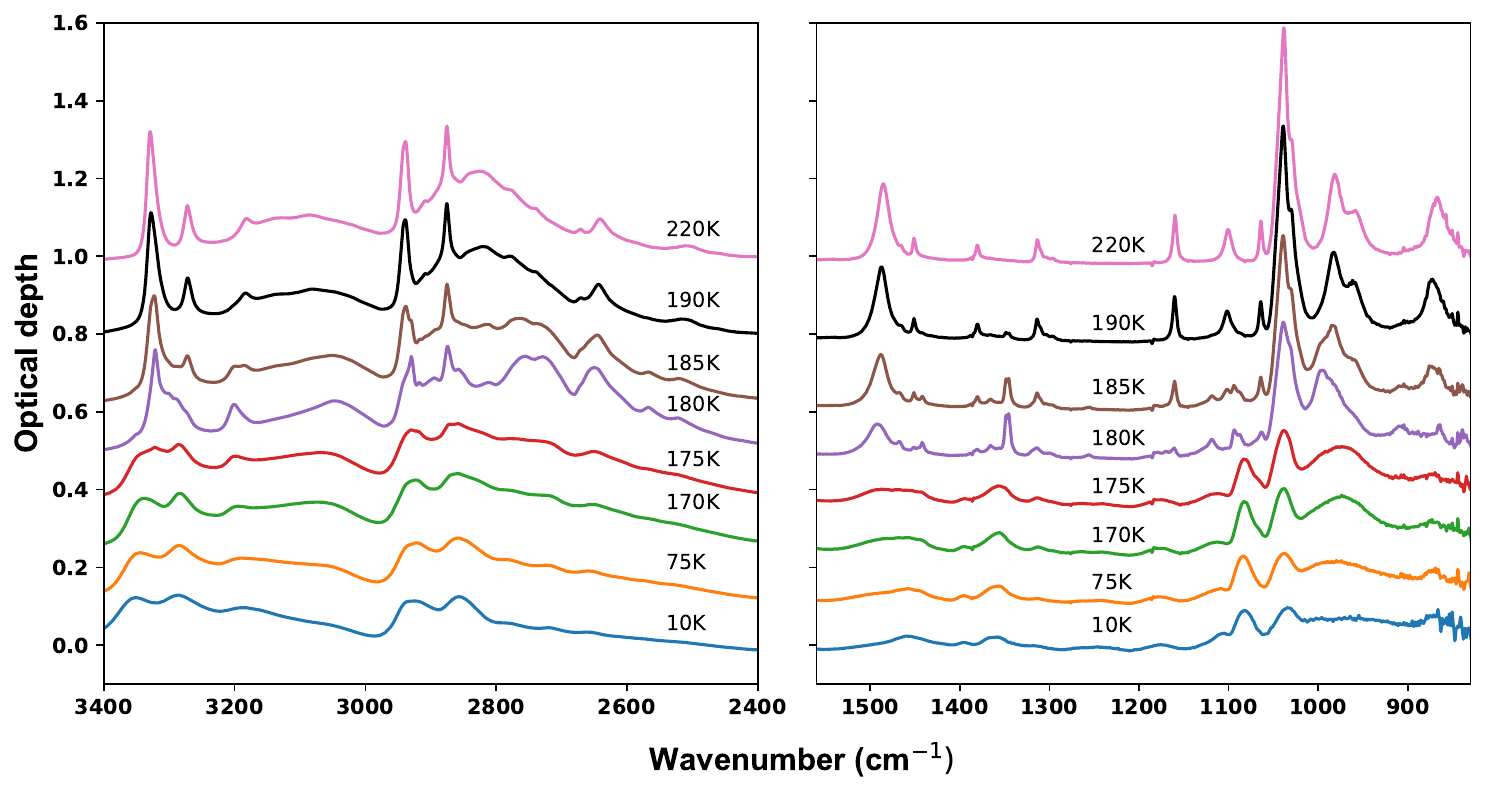}}
\caption{A series of mid-IR spectra of EA is shown at different temperatures and two different spectral regions. About 200 ML of EA were initially condensed on the KBr surface at 10 K and warmed up to 250 K with a linear heating ramp. Spectra are shifted in ordinate for clarity. The spectrum at 190 K is made bold to indicate the complete phase transition from amorphous to crystalline.}
   \label{fig2}
\end{center}
\end{figure*}

\begin{figure}
\begin{center}
\resizebox{\hsize}{!}{\includegraphics{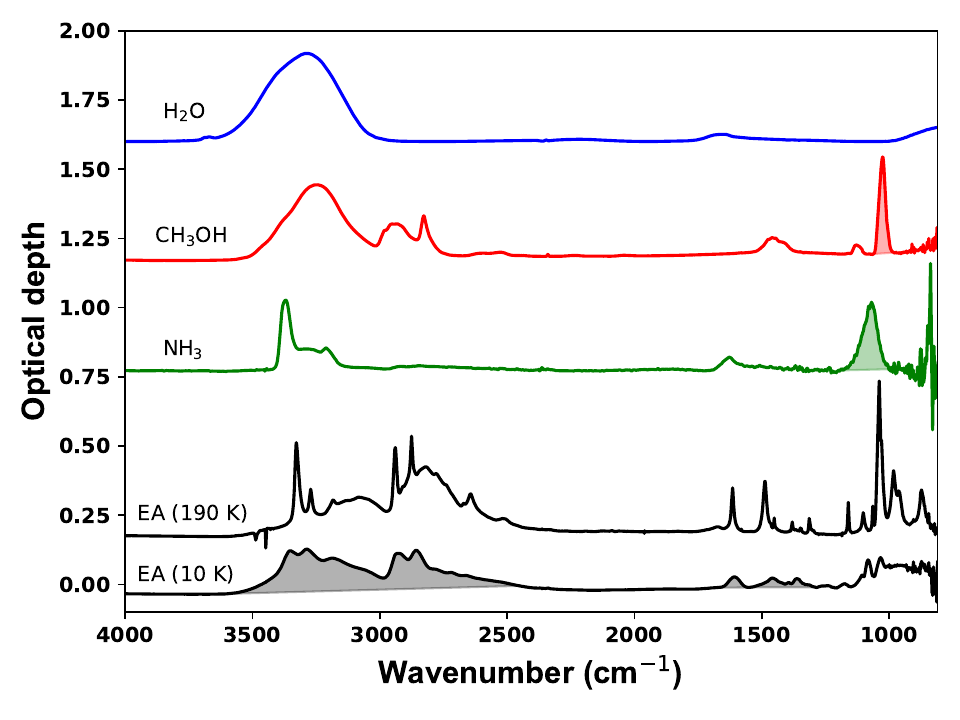}}
\caption{Mid-infrared spectra of pure EA and reference molecules NH$_3$ and CH$_3$OH at 10 K. The spectrum of H$_2$O (common interstellar ice component) at 15 K is taken from the Leiden Ice Database for Astrochemistry (LIDA). Bands selected for band strength estimation are indicated by the shaded areas.}
   \label{fig3}
\end{center}
\end{figure}

\section{Results and discussion}
\subsection{Infrared spectrum of EA ice}
Experimental IR spectra of EA in both its liquid and the gas phases have been reported in the literature \citep[and references therein]{silva1999}. Additionally, the adsorption behavior of EA on TiO$_2$ at temperatures of 35$^{\circ}$ and 275$^{\circ}$ C has been investigated \citep{tseng2010}. However, to date, a comprehensive study of EA ices at low temperatures has not been conducted. Matrix-isolated infrared spectra of EA at 13 K were measured in the laboratory, focusing only on the spectral regions between 3800-3300 and 600-200 cm$^{-1}$ \citep{rasanen1983}. Figure~\ref{fig1} illustrates the IR spectrum of EA deposited on a blank KBr substrate at 10 K. To aid in the assignment of the fundamental vibration modes associated with this spectrum, a fully anharmonic infrared spectrum of a single EA molecule was computed. The computed infrared spectrum is also shown in Figure~\ref{fig1}. Since the formation of ice significantly alters the spectrum of EA, especially in the stretching vibration region, the assignment of absorption bands is also based on previous laboratory data and is shown in Table ~\ref{table:1}. Although the computed spectrum agrees well with the experiment in the fingerprint region (i.\,e. low wavenumber side), there are a few discrepancies in the higher wavenumber side. Notably, the O-H stretching band of EA in the experiment deviates significantly from results obtained using computations. The possible cause of deviation is the presence of an extended H-bond network that exists in EA ice which was not considered for free hydroxy and amino groups in the computation.

\subsection{Thermal variation of Infrared spectrum of EA ice}
A pure EA ice grown at 10 K was heated with a continuous linear ramp of 10 K min$^{-1}$ up to 140 K. The spectral changes were monitored with FTIR. After 140 K, the sample was heated in increments of 5 K with the same ramp. After reaching each heating step, heating was paused for an IR measurement and then resumed until a final temperature of 250 K was reached. While the phase transition was observed between 175 and 185 K, EA sublimation started at 220 K, well above the water sublimation temperature, and continued until 240 K. The reproducibility was confirmed through a subsequent experiment. It is worth mentioning at this point that the maximum desorption temperature for the EA multilayer (about 200 ML) is 40 K higher than that measured for small coverages (a few MLs) by the TPD method described in the next section. Figure~\ref{fig2} shows the spectra at different temperatures during the annealing process of EA.

Features of interest are mostly located within these two ranges: 3600 -- 2500 cm$^{-1}$ and 1600 -- 800 cm$^{-1}$. For the higher wavenumber range, the band observed at 3351 cm$^{-1}$, associated with asymmetric NH$_2$ stretching, sharpens with increasing temperature, and the peak is eventually red-shifted to 3338 cm$^{-1}$ at 145 K. The band then remains unchanged until 170 K. During the phase transition, the 3338 cm$^{-1}$ band begins to disappear and is replaced by a sharp 3322 cm$^{-1}$ band. With further temperature increase, the 3322 cm$^{-1}$ band blue-shifts to 3327 cm$^{-1}$ until the complete desorption of EA. Meanwhile, the band at 3286 cm$^{-1}$, ascribed to symmetric NH$_2$ stretching, grows sharper until the sample reaches 175 K but without a change in the peak position. During the phase transition, the 3286 cm$^{-1}$ vanishes and a new band centered at 3271 cm$^{-1}$ appears. The band assigned to OH stretching at 3185 cm$^{-1}$ blue-shifts to 3200 cm$^{-1}$ and grows sharper up to 180 K. Beyond that the 3200 peak starts to disappear and a new band is formed at 3182 cm$^{-1}$, which is closer to the original band but better resolved. A shoulder peak at 3055 cm$^{-1}$ on the lower wavenumber side of the OH stretching band becomes prominent while heating the ice beyond 100 K. The peak then remains unchanged until 175 K before broadening. For the asymmetric CH$_2$ stretching band at 2928 cm$^{-1}$, although no change in the peak position was observed, the peak becomes sharper by heating up to 175 K. After this, the peak starts to disappear and instead, a new band is formed with a peak centered at 2939 cm$^{-1}$. On the other hand, a symmetric CH$_2$ stretch band at 2857 cm$^{-1}$ shifts up to 2864 cm$^{-1}$ at 170 K. The peak then disappears and also forms a new band with a peak at 2875 cm$^{-1}$. Apart from the bands mentioned in Table~\ref{table:1}, there were a few small peaks centered around 2777 cm$^{-1}$, 2718 cm$^{-1}$ and 2653 cm$^{-1}$ observed at 10 K which remained at the same position even after annealing until 170 K. All these peaks resolve upon annealing and show a considerable shift of up to 15 cm$^{-1}$ beyond 180 K. Two new peaks appear in EA ice spectrum beyond 180 K at 2565 cm$^{-1}$ and 2522 cm$^{-1}$. Close to the desorption temperature, the 2565 cm$^{-1}$ disappears while the 2522 cm$^{-1}$ peak shifts to 2510 cm$^{-1}$.

Simultaneously, for the lower wavenumber range, the 1607 cm$^{-1}$ band assigned to NH$_2$ bending shows no change in peak position until about 175 K. Between 175 and 185 K, 1607 peak disappears and instead, two new features appear -- a strong band at 1614 cm$^{-1}$ and a weak band at 1668 cm$^{-1}$. Similarly, the 1458 cm$^{-1}$ peak assigned to CH$_2$ bending gradually smears out until 175 K, and beyond that, during the phase transition, it forms a strong band at 1486 cm$^{-1}$ and a weak band at 1450 cm$^{-1}$. We also notice a peak emerging at 1467 cm$^{-1}$ which eventually becomes a shoulder to the 1486 cm$^{-1}$ peak. The 1395 cm$^{-1}$ band shows almost no change in the peak position until 175 K, while between 175 and 185 K, the peak shifts to 1380 cm$^{-1}$. On the other hand, 1360 cm$^{-1}$ peak becomes sharp but remain around 1357 cm$^{-1}$ until 175 K; beyond that the peak splits into a strong 1346 cm$^{-1}$ and a weak 1366 cm$^{-1}$. After 185 K, both the peaks start to disappear gradually. The 1253, 1175, 1082, and 1033 cm$^{-1}$ bands show no change in the peak position until 175 K. During the phase transition, we see peak shifting to 1160 cm$^{-1}$ in the case of 1175 cm$^{-1}$ band while 1082 and 1033 cm$^{-1}$ bands show splitting to 1100 and 1063 cm$^{-1}$ peaks for the former and 1038 and 1030 cm$^{-1}$ peaks for the latter. Two new peaks at 970 and 1313 cm$^{-1}$ start to appear as we start to heat the EA ice and remain unchanged until 175 K. During the phase transition, 970 cm$^{-1}$ peak splits into 982 and 960 cm$^{-1}$ peaks. After 185 K, a new peak appears at 870 cm$^{-1}$.

\subsection{Band strength estimation}

In the laboratory, optical laser interferometry is frequently paired with complementary techniques to determine the infrared band strengths of solid molecules \citep[and references therein]{bouilloud2015, hudson2022}. The precise estimation of absolute band strengths relies on prior knowledge of the refractive index and density of the molecule under study, which laser techniques can help with. In the present work, we have determined the infrared band strengths of EA by three different methods, none of which involve laser interferometry. The band strengths derived by the first two methods require one to compare the absorption spectra of molecules with known reference band strengths while for the last method, one doesn't require such a comparison.

Infrared band strength ($A$, cm molecule$^{-1}$) of any absorption band of a molecule is given by:

\begin{equation}
A=  \frac{1}{N} \int \tau (\nu) d\nu
\label{Eq1}
\end{equation}

where N is the column density of the molecule in molecules cm$^{-2}$, $\tau$ the optical depth of the absorption band (unitless), and $\nu$ the wavenumber frequency in cm$^{-1}$. To determine the band strength, the column density of the molecule has to be measured. In the case of the first method, we considered that the column density will depend on the gas pressure difference ($\Delta p$) measured in the gas inlet system during the molecule deposition. The pressure difference is measured via the capacitance diaphragm vacuum gauge (Pfeiffer Vacuum, CCR 364). The gas pressure obtained this way is independent of the gas type and the concentration. Therefore, assuming an ideal gas equation and considering that all the other parameters remain unchanged from one measurement to the other, $N$ would be proportional to $\Delta p$ in the gas inlet system. A similar dependence has been shown previously \citep{gonzalez2022}.
% N was defined as the column density of the molecule in molecules cm−2. It will depend on many factors not only on delta P.
For this method of band strength estimation of EA, we use a reference molecule and re-arrange Equation~\ref{Eq1} as the following:

\begin{equation}
A_{EA} = \frac{\Delta p_{ref}\int \tau_{EA} (\nu) d\nu}{\Delta p_{EA} \int \tau_{ref} (\nu) d\nu} A_{ref}
\label{Eq2}
\end{equation}

%The column density of the molecule can be estimated from the Hertz–Knudsen formula that describes the non-dissociative adsorption of a gas molecule on a surface under the assumption that the sticking coefficient of the gas molecules on the surface is unity:

%\begin{equation}
%N=  \frac{p}{\sqrt{2\pi mk_BT}}t
%\label{Eq2}
%\end{equation}

%where, $p$ is the gas pressure in the vacuum chamber, $m$ the mass of the molecule, $k_B$ the Boltzmann constant, $T$ the gas temperature in Kelvin (300 K) and $t$ the dosing time.

The integrated band areas for EA and the reference molecules were evaluated from the infrared spectra. We have used CH$_3$OH and NH$_3$ as reference molecules. Bands of CH$_3$OH and NH$_3$ selected for the integrated optical depth estimations are shown with shaded regions in Figure~\ref{fig3} and the respective absolute band strengths are listed in Table~\ref{table:3}. In the case of NH$_3$'s 1071 cm$^{-1}$ peak, the absolute band strength was calculated using the imaginary part of the optical constants k($\nu$) given as a function of wavenumbers that was obtained from NASA Cosmic Ice Laboratory database\footnote{https://science.gsfc.nasa.gov/691/cosmicice/constants.html}. The optical constants in turn allow one to determine the absolute absorption coefficient $\alpha$ for each wavenumber using the relation $\alpha$ = 4$\pi\nu$k and generate a computed IR spectrum. The integration of the 1071 cm$^{-1}$ peak can then be used to eventually derive the absolute band strengths using the following equation \citep{hudson2014}:

\begin{equation}
\label{Eq3} 
A = \frac{M}{\rho N_\mathrm{A}} \int_{band} \alpha (\nu) d\nu
\end{equation}

where M is the molar mass, N$_\mathrm{A}$ the Avogadro constant, and $\rho$ the density of the molecule. $\rho_\mathrm{NH_3}$ = 0.68 g cm$^{-3}$ was used for the calculation \citep{hudson2022} which ultimately yields A$_\mathrm{NH_3}$ = 22.3 $\times$ 10$^{-18}$ cm mol$^{-1}$. 

To test how well this method works, we calculated the absolute band strength of NH$_3$ using CH$_3$OH as a reference and vice versa. In both cases, a set of two different $\Delta p$ values were taken into consideration for each molecule. In the former case, A$_\mathrm{NH_3}$ = 26$\pm$6 $\times$ 10$^{-18}$ cm mol$^{-1}$ while for the latter, A$_\mathrm{CH_3OH}$ = 15$\pm$3 $\times$ 10$^{-18}$ cm mol$^{-1}$ was obtained. Within the errors, the deviation is around 10\%. The estimated values of EA band strength for the three different integrated ranges by the two reference molecules are well within errors and they are summarized in Table~\ref{table:2}. One standard deviation was used for the error calculation.

The second method employed QMS measurements during TPD to obtain the column density of the molecules desorbed from the surface for band strength estimation. The species used were monitored through their primary mass fragment: $m/z$ = 31 (CH$_3$OH) and $m/z$ = 30 (EA). It is worth mentioning that NH$_3$ was excluded as a reference molecule in the second method due to the uncertainties associated with the relative transmission efficiency of the QMS between the m/z = 30 (EA) and m/z = 17 (NH$_3$). Such uncertainty can be neglected when using CH$_3$OH as the reference molecule for Method 2 since the m/z ratios monitored are similar (m/z = 31 for CH$_3$OH and m/z = 30 for EA). The integrated QMS signal measured during the TPD of a molecule is proportional to $N$ and depends on several other parameters as highlighted by \citet{mate2023}. The band strength of EA in comparison to CH$_3$OH can then be expressed as:

\begin{equation}
\label{Eq4}    
A_{EA} = \frac{I_{ref}\sigma_{EA}f_{EA}\int \tau_{EA} (\nu) d\nu}{I_{EA}\sigma_{ref}f_{ref}\int \tau_{ref} (\nu) d\nu} A_{ref}
\end{equation}

where I is the integrated QMS signal, $\sigma$ the ionization cross-section for the first ionization of the species at 70 eV electron energy employed in the QMS, and $f$ the fragmentation factor, which is the fraction of the primary fragment mass with respect to all the other fragment masses. The $\sigma$ values for the two species are $\sigma_\mathrm{CH_3OH}$ = 4.49 $\AA^2$ \citep{hudson2003} and $\sigma_\mathrm{EA}$ = 13.255 $\AA^2$ based on complex scattering potential-ionization contribution (CSP-ic) method \citep{chakraborty2024}. The f values are f$_\mathrm{CH_3OH}$ = 0.404,and f$_\mathrm{EA}$ = 0.648 (NIST database). The transmission efficiency can be considered equal. Considering all these correction parameters, the estimated EA band strength values deviate significantly from the values obtained by Method 1 as seen in Table~\ref{table:2}. The possible cause for such deviation could be due to the uncertainty in the determination of different parameters used in Equation~\ref{Eq4}.

%, $\sigma_{NH_3}$ = 3.036 $\AA^2$ (extracted from the online database of the National Institute of Standard and Technologies (NIST)) $f_{NH_3}$ = 0.526 

\begin{figure}
\begin{center}
\resizebox{\hsize}{!}{\includegraphics{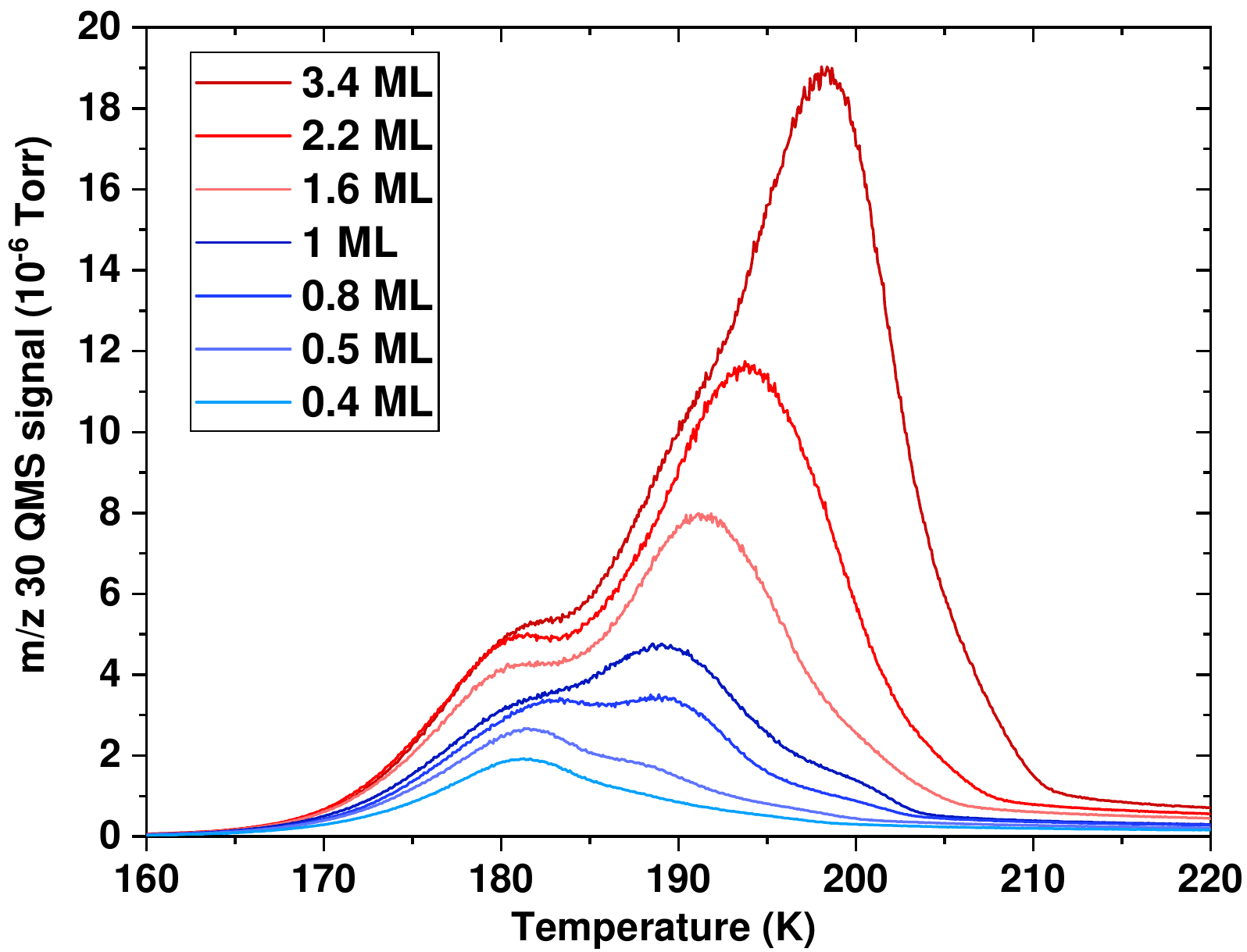}}
\caption{Thermal desorption signal of EA from pure EA ices deposited at 100 K, and using a ramp of 10 K min$^{-1}$. The coverage associated with each curve is based on the considerations as explained in the text.}
   \label{fig_tpd}
\end{center}
\end{figure}

The last method that we used (Method 3) consists of a series of Temperature Programmed Desorption (TPD) measurements, which is a well-known technique in laboratory astrophysics \cite[e.g.][]{Burke_2010, Doronin_2015}. EA ices are grown at 100 K by exposing a KBr substrate to a partial pressure between 10$^{-9}$ and 10$^{-8}$ Torr of EA for a few minutes. Then the substrate is warmed up to 250 K using a ramp of 10 K min$^{-1}$ and the thermal desorption of EA is probed as a function of the substrate temperature by following the $m/z$ = 30 channel of the QMS. Figure \ref{fig_tpd} shows the TPD curves for different depositions of EA ices. According to the Polanyi-Wigner description of thermal desorption, we estimate the monolayer (ML) coverage of EA from these TPD curves. For multi-layer depositions (red curves), the TPD curves display a common leading edge at the start of the desorption kinetics, from 165 K to 175 K, whereas for monolayer or sub-monolayer depositions (blue curves), this leading edge differs for each deposition. Based on those considerations and assuming that depositions are homogeneous and that the TPD integral is linear with the coverage, we estimate the EA coverage as shown in the legend of Figure \ref{fig_tpd}. Infrared measurements are also taken after each deposition before the onset of the TPD measurements. A good signal-to-noise ratio could only be obtained in the IR spectrum for coverages greater than 3.4 ML. Therefore, we used that in the determination of the infrared band strength. Knowing the ice thickness associated with this IR spectrum, we therefore deduced the band strengths as displayed in Table \ref{table:2} (Method 3). The evaluated band strength values show less deviation than Method 2. In Method 3, there could be uncertainties in the accurate determination of monolayer coverage, which could explain the deviation between Method 1 and Method 3. The error in Method 3 is computed from the relative difference obtained on the coverage of the 3.4 ML TPD curve when considering the actual 1.6 ML TPD curve to correspond to 1 ML. This results in a 38\% uncertainty. Note that we have only used a few monolayer coverage equivalent of EA on the cold surface for band strength estimation by the three methods described above.

The advantage of Method 1 is the ease of band strength measurements of unknown molecules without the need for a dedicated experimental setup and the time for undertaking the experiment. This is especially suitable for larger organic molecules that are unstable and can undergo polymerization with time. The major advantage of Method 3 is that one can easily obtain the binding energy of the molecule of interest together with the band strength. On the other hand, the inconsistent variation of band strength estimation with Method 2 highlights the need for measuring parameters such as the ionization cross-section, the fragmentation factor, and the transmission efficiency of molecules in the laboratory through the QMS. Many researchers use an alternative approach, which involves laser interferometry. However, this method requires additional equipment and knowledge of the refractive index mass density of the ice formed. Furthermore, it assumes that the ice layer on the substrate is perfectly flat with zero porosity, which is not always the case. Therefore, this method can only be recommended when Method 1 cannot be applied, such as when measuring substances with low vapor pressures. Based on the discussion in the paper we think that the three measurement approaches taken in this paper can all be used in ways described in the paper. From the comparison of the band strength values obtained from three methods, we think the direct method based on mass spectrometer and indirect method based on pressure gauge would provide a reliable determination of the band strengths.

Furthermore, alongside the three aforementioned methods, we have included absolute infrared band strengths derived from theoretical calculations in Table~\ref{table:1}. A satisfactory agreement exists between experimental and theoretical values for the NH$_2$ bending mode at 1607 cm$^{-1}$, considering the inherent approximations in both experimental and theoretical approaches. While this method may be effective for estimating individual vibration modes, it may not be suitable when considering combinations of various modes.

The observed double peak structure in the TPD could arise from two potential causes. Firstly, it may result from some degree of non-uniformity in film growth across the substrate during EA deposition using a 0.5 mm diameter capillary tube \citep{potapov2019}. Secondly, the presence of minor hydroxyl group contamination on KBr could induce H-bonding interactions with EA, thereby contributing to the observed double peak structure. For multilayer desorption, with the coincident leading edge behaviour seen in the TPD trace, the desorption energy was then determined using an Arrhenius analysis of the leading edge region for the 3.4 ML curve as described previously in details \citep{suhasaria2015}. The method yields the multilayer desorption energy of 0.61$\pm$0.01 eV and the pre-exponential factor of $0.5\times 10^{27}$ molecule cm$^{-2}$ s$^{-1}$. In the literature, the binding energy of EA from carbon and water surface has been predicted via machine learning to be around 0.7$\pm$0.2 eV \citep{villadsen2022}. Although one cannot compare the values directly, the binding energy of EA from pure EA ice seems to be in good agreement with the predicted values within the errors. For systems with more complex desorption kinetics, transition state theory can be employed as a method to derive the pre-exponential factor \citep{thrower2013, suhasaria2015}.

\begin{figure}
\begin{center}
\resizebox{\hsize}{!}{\includegraphics{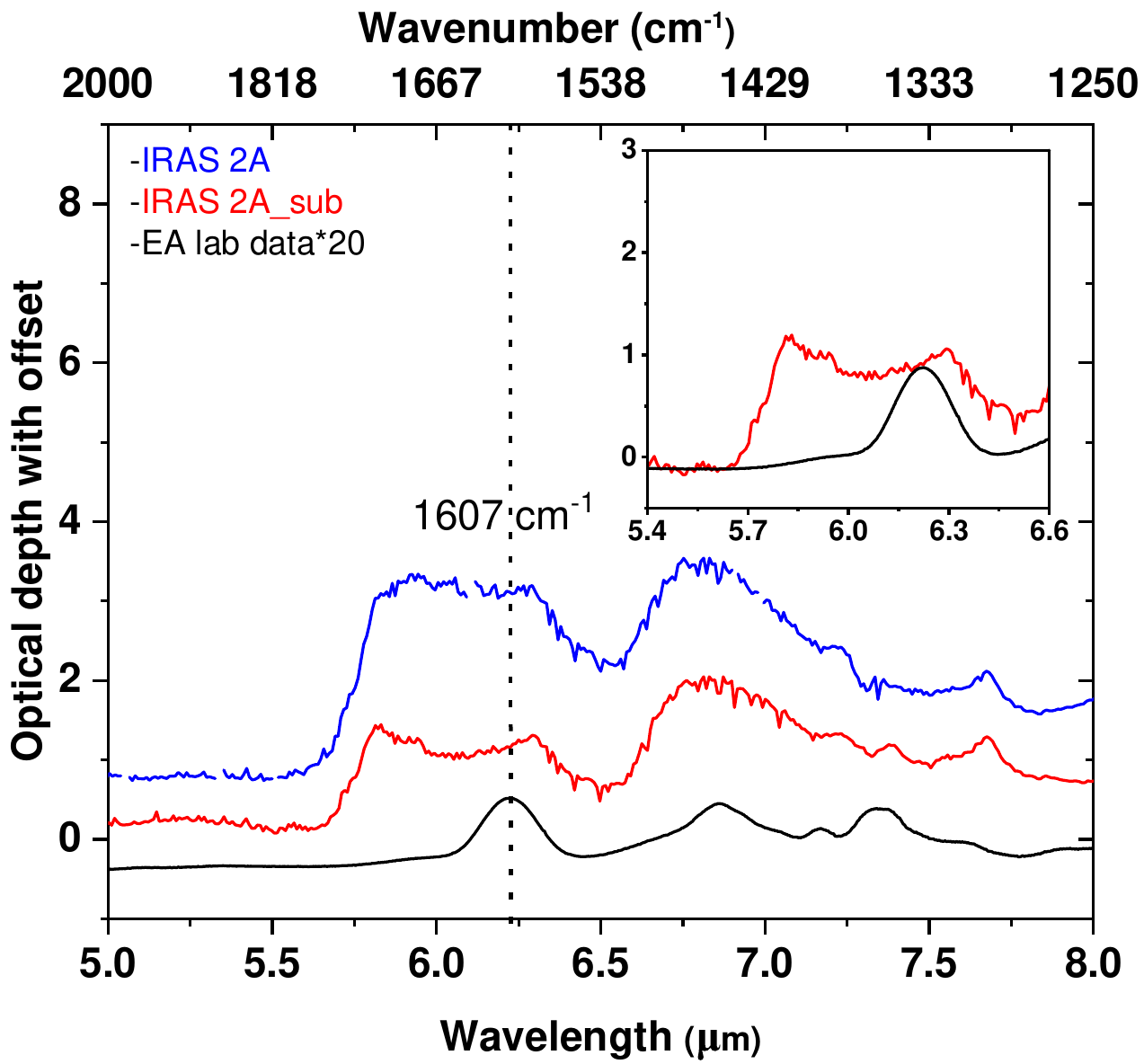}}
\caption{Comparison of the JWST/MIRI MRS spectra of IRAS 2A (blue curve) and IRAS 2A with water:silicate features subtracted (red curve) to the laboratory-measured EA ice profile at 10 K. For clearer comparison, all spectra are vertically shifted, with the laboratory spectrum intensity is multiplied by 20, and the EA band at 1607 cm$^{-1}$ is indicated by dashed line. The inset provides a closer examination of the water:silicate subtracted IRAS 2A spectrum alongside the fitted EA profile for upper limit estimation.}
   \label{fig5}
\end{center}
\end{figure}

\begin{table*}
\caption{IR band positions of EA from experiment and calculation.}              % title of Table
\label{table:1}      % is used to refer this table in the text
\centering                                      % used for centering table
\begin{tabular}{c c c c }          % centered columns (4 columns)
\hline\hline                        % inserts double horizontal lines
Experimental & Calculated & Assignment & Calculated A'\\ 
% table heading
\hline                                   % inserts single horizontal line
3351 & 3656&$\nu_{1}$ (asymm. NH$_2$)&$3.47\times10^{-18}$ \\  % inserting body of the table
3286 &3438& $\nu_{2}$ (symm. NH$_2$)&$1.20\times10^{-18}$\\
3185 && $\nu_{3}$ (OH)&\\
2928 &2958&$\nu_{4}$ (asymm. CH$_2$)&$16.75\times10^{-18}$\\
2857 &2888&$\nu_{5}$ (symm. CH$_2$)&$9.64\times10^{-18}$\\
1607 &1580&$\delta_{6}$ (NH$_2$)&$8.31\times10^{-18}$\\
1458 &1460& $\delta_{7}$ (CH$_2$)&$0.81\times10^{-18}$\\
1395&1380&$\nu_{8}$ (CC)+ $\delta_{9}$ (OH, CH)&$2.58\times10^{-18}$\\
1360&1356&$\tau_{10}$ (NH, OH)&$2.95\times10^{-18}$\\
1253&1230&Skeletal vibrations* &$0.90\times10^{-18}$\\
1175 &1138&Skeletal vibrations* &$2.53\times10^{-18}$\\
1082&1090&$\nu_{11}$ (CN, CO) +$\delta_{12}$ (OH)+ $\tau_{13}$ (CC) &$6.97\times10^{-18}$\\
1033&976&$\nu_{14}$ (CO)+ $\delta_{15}$ (NH, CH)+ $\tau_{16}$ (CC)& $6.63\times10^{-18}$ \\
\hline                                             %inserts single line
\end{tabular}
\tablefoot{Mode assignments in the lower wavenumbers are based on the calculation. The calculated band strengths (A', cm molecules$^{-1}$) have been added in the last column. Vibration modes have been denoted by Greek letters: stretching ($\nu$), bending ($\delta$) and twisting ($\tau$). Assignments marked with asterisks (*) arise due to strong coupling between stretching and bending modes of different atoms in the EA.}
\end{table*}

\begin{table*}
\caption{Estimated IR band strengths of EA ice at 10 K using three different methods.} 
\label{table:2}      % is used to refer this table in the text
\centering                                      % used for centering table
\begin{tabular}{c c c c c c} 
\hline\hline    % inserts double horizontal lines
\multirow{3}{*}{Mode} &
\multirow{2}{*}{Integrated region} & 
\multicolumn{2}{c}{{Method 1}} & 
Method 2 &
Method 3 \\
& & NH$_3$ & CH$_3$OH & CH$_3$OH & EA \\ 
&cm$^{-1}$& \multicolumn{4}{c}{$\times10^{-18}$ cm molecules$^{-1}$} \\
\hline                                   % inserts single horizontal line
$\nu_{1}$+$\nu_{2}$+$\nu_{3}$+$\nu_{4}$+$\nu_{5}$ & 3650 -- 2450& 463 $\pm$ 65 & 528 $\pm$ 137 & 202 $\pm$ 61 &386 $\pm$ 145 \\
$\delta_{6}$ & 1690 -- 1550 & 10.2 $\pm$ 0.3 & 12 $\pm$ 3 &5 $\pm$ 2 &7.8 $\pm$ 2.9\\
$\delta_{7}$+$\nu_{8}$+$\delta_{9}$+$\tau_{10}$&1550 -- 1285 & 17.3 $\pm$ 0.4 & 20 $\pm$ 4 &9 $\pm$ 3 &16 $\pm$ 6 \\
\hline
\end{tabular}
\tablefoot{Note that there are large uncertainties in the determination of band strengths by Method 2. The possible reasons are discussed in the text.}
\end{table*}

\begin{table*}
\caption{IR band strengths for reference and molecules with functional group similar to EA}       
\label{table:3}     
\centering                                      
\begin{tabular}{c c c c c}         
\hline\hline                        % inserts double horizontal lines
Molecule & Band position & Assignment & Integrated region & Band strengths \\ 
 & cm$^{-1}$ && cm$^{-1}$ & cm molecule$^{-1}$ \\% table heading
\hline                                   % inserts single horizontal line
NH$_3$$^{\textit{a}}$ & 1071 & umbrella & 1300 -- 950 & $22.3\times10^{-18}$ \\  % inserting body of the table
CH$_3$OH$^{\textit{b}}$ & 1030 & $\nu$ (CO) & 1066 -- 967 & $16.8\times10^{-18}$ \\
CH$_3$NH$_2$(Methylamine)$^{\textit{c}}$ & 1600 & $\delta$ (NH$_2$)& & $4.3\times10^{-18}$ \\
CH$_3$NHCH$_2$OH(N-methylaminomethanol)$^{\textit{d}}$ & 1414 & $\delta$(CH)+$\delta$(NH) & & $6\times10^{-19}$ \\
CH$_3$CH=NH(ethanimine)$^{\textit{e}}$ & 1650 & $\nu$ (C=N)& & $4.4\times10^{-18}$ \\
CH$_2$=C=NH(Ketenimine)$^{\textit{f}}$ &  & & & $7.2\times10^{-17}$ \\
NH$_2$CH(CH$_3$)OH($\alpha$-aminoethanol)$^{\textit{g}}$ & 1413 & $\delta$ (CH) && $1.4\times10^{-18}$ \\
CH$_3$CH$_2$OH(ethanol)$^{\textit{h}}$ & 1050 & $\nu$ (CO) & 1067 -- 1011 & $14.1\times10^{-18}$\\
\hline                                             %inserts single line
\end{tabular}
\tablefoot{$^{\textit{a}}$ We have calculated the absolute band strength of ammonia based on the apparent value estimated by \citet{hudson2022}, $^{\textit{b}}$ \cite{gerakines2020}, $^{\textit{c}}$ \cite{holtom2005}, $^{\textit{d}}$ \cite{vinogradoff2013}, $^{\textit{e}}$ \cite{canta2023}, $^{\textit{f}}$ \cite{hudson2004}, $^{\textit{g}}$ \cite{duvernay2010} and $^{\textit{h}}$ \cite{hudson2017}}
\end{table*}

\section{Astrophysical implications}

EA has been identified in both the interstellar medium and meteorites \citep{rivilla2021, glavin2010}, indicating its likely origin in space and its potential delivery to early Earth. This suggests a possible role for EA in the formation of crucial prebiotic molecules like amino acids, potentially contributing to the emergence of life \citep{rivilla2021}. Furthermore, since EA can be formed via aminoketene, a link between EA and peptides, which are also produced through aminoketene, may exist \citep{Krasnokutski2022}. Therefore, the detection of EA in both gas and solid phases across various sources such as molecular clouds and protoplanetary disks could indicate chemical processes essential for the origin of life. Accurately determining EA abundance or providing at least an upper limit in different conditions is therefore crucial for gaining reliable insights.

The gas-phase molecular column density of EA toward the G+0.693 molecular cloud in the SgrB2 complex in the Galactic Center was estimated at $1.5\times10^{13}$ cm$^{-2}$. This suggests an estimated abundance ranging from 0.9 to $1.4\times10^{-10}$ relative to molecular hydrogen and the EA/H$_2$O abundance ratio of the order of 10$^{-6}$ \citep{rivilla2021}. In the Almahata Sitta meteorite, the ratio EA/H$_2$O was also found to be of the same order \citep{glavin2010}. While ice signatures of EA have not been detected, it is reasonable to assume that the EA abundance is significantly higher in the solid state than the gas phase given the various solid-state pathways suggested on grains \citep{rivilla2021}. In the gas phase, EA molecules might easily dissociate under UV photon irradiation due to their predominantly single bonds and relatively small size, making them more fragile and less capable of dissipating energy. Conversely, higher stability and potentially consequential reactions are expected for EA in the solid state.

The enhanced sensitivity of the Medium Resolution Spectrograph (MRS) \citep{wells2015} of the Mid-InfraRed Instrument (MIRI) \citep{rieke2015, wright2015} onboard the JWST compared to its predecessors raises the possibility of identifying solid EA in various astrophysical environments in the future. Figure~\ref{fig5} illustrates a comparison of the pure EA ice spectrum at 10 K with an observed JWST spectra of the low mass protostar IRAS 2A taken with the MRS MIRI \citep{rocha2024}. As can be seen from Figure~\ref{fig5}, the NH$_2$ deformation mode of EA coincides with the strong features observed in the JWST spectra. However, many other molecules with NH$_2$ and even C=C bands can contribute to the same feature. Figure~\ref{fig3} shows the spectrum of pure amorphous and crystalline EA ice along with the spectra of some of the common polar interstellar ice components. All of the EA ice features overlap with the common interstellar ices and it could be challenging to significantly constrain the EA abundance. However, within environments where grain temperatures exceed the water desorption temperature ($T_d$ $\sim$ 160 K), the majority of volatile components within the ice mantle will sublimate. This process would result in an elevated concentration of EA ($T_d$ > 180 K) in the solid state. In this case, the band at 1607 cm$^{-1}$ assigned to NH$_2$ bending can be used to determine the column density of EA. Moreover, within astrophysical environments characterized by elevated temperatures (> 170 K), such as those found in protoplanetary disks, EA ice may manifest in a crystalline state. This crystalline form displays distinct sharp characteristics that distinguish it from other types of refractory ice. Finally, even at low temperatures (< 150 K) where volatile ice exists, it is possible to subtract the spectral features of known components such as NH$_3$, CH$_3$OH, H$_2$O, and even silicates to detect EA. This process may result in a loss of sensitivity, but it will not prevent the detection or estimation of the upper limit of the abundance of EA.

In this work, we have estimated the upper limit of EA by comparing the laboratory data with the observed spectrum. The spectrum from IRAS 2A with water:silicate features subtracted was used to constrain the EA upper limit abundance. The EA feature at 1607 cm$^{-1}$ (6.22 \textmu m) marked with vertical dashed line was used for this purpose. The column density upper limit ($N$, in cm$^{-2}$) of EA can be estimated using the following relation described in detail in \citet{rachid2021}:

\begin{equation}
\label{Eq5}    
N_{EA} \leq \frac{\tau_\nu \times FWHM}{A}
\end{equation}

where, $A$ = $1\times10^{-17}$ cm molecule$^{-1}$ is the band strength and FWHM = 50.5 cm$^{-1}$ for the 1607 cm$^{-1}$ EA feature. $\tau_{\nu}$ is the maximum optical depth of the water:silicate subtracted peak where the EA peak centered at 1607 cm$^{-1}$ fits. The inset to Figure~\ref{fig5} shows the procedure for obtaining the maximum optical depth. Using this method, an upper limit of $\leq$ $4.4\times10^{18}$ cm$^{-2}$ was estimated, which corresponds to about 15\% with respect to H$_2$O, assuming a column density of $3\times10^{19}$ cm$^{-2}$ for water \citep{rocha2024}. An abundance of EA this high seems improbable, and as previously noted, the 1607 cm$^{-1}$ EA feature used for estimating abundance could also originate from other molecules. Apart from the upper limit estimation, the band strength estimation is also needed for the quantitative estimation of the photolysis products of EA. Improving the accuracy with which we know the initial concentration of the starting material will help us to better calculate the product yield.

 In Table~\ref{table:3} we have listed the band strengths of the reference molecules (NH$_3$ and CH$_3$OH) used for band strength estimation of EA and of some other molecules that could potentially exist in interstellar medium and have one or both the functional groups (saturated and unsaturated) present in EA. Although one cannot directly compare the band strength values, a closer look indicates that the order of magnitude is similar to other molecules listed.

\section{Conclusions}

Solid EA was deposited on KBr substrate at 10 K and infrared spectra of the amorphous ice have been measured. Band assignments have been made by comparison with the calculated spectra. The EA ice was then annealed up to 250 K. Transmission spectra were recorded over the entire temperature range. Between 175 and 185 K, the amorphous EA phase undergoes a transition to the crystalline phase. Infrared band strengths were measured for amorphous EA ice. Three different methods were used to derive the band strengths. One of the three methods used a complete set of TPDs for EA from submonolayer coverage to the multilayer regime. The multilayer binding energy and the pre-exponential factor were determined experimentally. For those gases whose pressure can be precisely measured in the gas line, the first tested method of determining absorption band strengths seems to be the simplest, giving satisfactory measurement accuracy, and can be recommended for use in laboratories. The second method requires precise knowledge of ion transport in the mass spectrometer, which is often missed and the third method provides more additional information but is relatively time-consuming and has also uncertainties. A comparison between the laboratory data of pure EA and the the JWST MIRI spectrum of the low-mass protostar IRAS 2A enabled us to derive an upper-limit for the EA ice abundances with respect to solid H$_2$O.

\begin{acknowledgements}

The authors are grateful to W.R.M. Rocha and E.F. van Dishoeck for providing the MIRI MRS JWST data. This work has been supported by the European Research Council under the Horizon 2020 Framework Program via the ERC Advanced Grant Origins 83 24 28. We would also like to acknowledge funding through the DeutscheForschungsgemeinschaft (DFG) project No. JA 2107/11-1.
S.A.K. is grateful for the support from DFG (grant no. KR 3995/4-2). The authors also acknowledge support from the NanoSpace COST action (CA21126 – European Cooperation in Science and Technology).

\end{acknowledgements}

\bibliographystyle{aa}
\bibliography{Reference}

\begin{thebibliography}{37}
\expandafter\ifx\csname natexlab\endcsname\relax\def\natexlab#1{#1}\fi

\bibitem[{Bernstein {et~al.}(2002)Bernstein, Dworkin, Sandford, Cooper, \& Allamandola}]{bernstein2002}
Bernstein, M.~P., Dworkin, J.~P., Sandford, S.~A., Cooper, G.~W., \& Allamandola, L.~J. 2002, Nat, 416, 401

\bibitem[{Bouilloud {et~al.}(2015)Bouilloud, Fray, B{\'e}nilan, Cottin, Gazeau, \& Jolly}]{bouilloud2015}
Bouilloud, M., Fray, N., B{\'e}nilan, Y., {et~al.} 2015, MNRAS, 451, 2145

\bibitem[{Burke \& Brown(2010)}]{Burke_2010}
Burke, D.~J. \& Brown, W.~A. 2010, Phys. Chem. Chem. Phys., 12, 5947

\bibitem[{Canta {et~al.}(2023)Canta, {\"O}berg, \& Rajappan}]{canta2023}
Canta, A., {\"O}berg, K.~I., \& Rajappan, M. 2023, ApJ, 953, 81

\bibitem[{Chakraborty {et~al.}(2024)Chakraborty, Sinha, \& Antony}]{chakraborty2024}
Chakraborty, I., Sinha, N., \& Antony, B. 2024, Radiat. Phys. Chem., 216, 111421

\bibitem[{Doronin {et~al.}(2015)Doronin, Bertin, Michaut, Philippe, \& Fillion}]{Doronin_2015}
Doronin, M., Bertin, M., Michaut, X., Philippe, L., \& Fillion, J.-H. 2015, J. Chem. Phys., 143, 084703

\bibitem[{Duvernay {et~al.}(2010)Duvernay, Dufauret, Danger, Theul{\'e}, Borget, \& Chiavassa}]{duvernay2010}
Duvernay, F., Dufauret, V., Danger, G., {et~al.} 2010, A\&A, 523, A79

\bibitem[{Frisch {et~al.}(2016)Frisch, Trucks, Schlegel, Scuseria, Robb, Cheeseman, Scalmani, Barone, Petersson, Nakatsuji, Li, Caricato, Marenich, Bloino, Janesko, Gomperts, Mennucci, Hratchian, Ortiz, Izmaylov, Sonnenberg, Williams-Young, Ding, Lipparini, Egidi, Goings, Peng, Petrone, Henderson, Ranasinghe, Zakrzewski, Gao, Rega, Zheng, Liang, Hada, Ehara, Toyota, Fukuda, Hasegawa, Ishida, Nakajima, Honda, Kitao, Nakai, Vreven, Throssell, Montgomery, Peralta, Ogliaro, Bearpark, Heyd, Brothers, Kudin, Staroverov, Keith, Kobayashi, Normand, Raghavachari, Rendell, Burant, Iyengar, Tomasi, Cossi, Millam, Klene, Adamo, Cammi, Ochterski, Martin, Morokuma, Farkas, Foresman, \& Fox}]{g16}
Frisch, M.~J., Trucks, G.~W., Schlegel, H.~B., {et~al.} 2016, Gaussian˜16 {R}evision {C}.01, gaussian Inc. Wallingford CT

\bibitem[{Gerakines \& Hudson(2020)}]{gerakines2020}
Gerakines, P.~A. \& Hudson, R.~L. 2020, ApJ, 901, 52

\bibitem[{Glavin {et~al.}(2010)Glavin, Aubrey, Callahan, Dworkin, Elsila, Parker, Bada, Jenniskens, \& Shaddad}]{glavin2010}
Glavin, D.~P., Aubrey, A.~D., Callahan, M.~P., {et~al.} 2010, Meteorit. Planet. Sci., 45, 1695

\bibitem[{Gonz{\'a}lez~D{\'\i}az {et~al.}(2022)Gonz{\'a}lez~D{\'\i}az, Carrascosa, Mu{\~n}oz~Caro, Satorre, \& Chen}]{gonzalez2022}
Gonz{\'a}lez~D{\'\i}az, C., Carrascosa, H., Mu{\~n}oz~Caro, G.~M., Satorre, M.~{\'A}., \& Chen, Y. 2022, MNRAS, 517, 5744

\bibitem[{Holtom {et~al.}(2005)Holtom, Bennett, Osamura, Mason, \& Kaiser}]{holtom2005}
Holtom, P.~D., Bennett, C.~J., Osamura, Y., Mason, N.~J., \& Kaiser, R.~I. 2005, ApJ, 626, 940

\bibitem[{Hudson {et~al.}(2003)Hudson, Hamilton, Vallance, \& Harland}]{hudson2003}
Hudson, J.~E., Hamilton, M.~L., Vallance, C., \& Harland, P.~W. 2003, Phys. Chem. Chem. Phys., 5, 3162

\bibitem[{Hudson \& Moore(2004)}]{hudson2004}
Hudson, R. \& Moore, M. 2004, Icarus, 172, 466

\bibitem[{{Hudson}(2017)}]{hudson2017}
{Hudson}, R.~L. 2017, Spectrochim. Acta A, 187, 82

\bibitem[{{Hudson} {et~al.}(2014){Hudson}, {Ferrante}, \& {Moore}}]{hudson2014}
{Hudson}, R.~L., {Ferrante}, R.~F., \& {Moore}, M.~H. 2014, \icarus, 228, 276

\bibitem[{Hudson {et~al.}(2022)Hudson, Gerakines, \& Yarnall}]{hudson2022}
Hudson, R.~L., Gerakines, P.~A., \& Yarnall, Y.~Y. 2022, ApJ, 925, 156

\bibitem[{Krasnokutski(2021)}]{Krasnokutski2021}
Krasnokutski, S.~A. 2021, Low Temp. Phys., 47, 199

\bibitem[{Krasnokutski {et~al.}(2022)Krasnokutski, Chuang, Jäger, Ueberschaar, \& Henning}]{Krasnokutski2022}
Krasnokutski, S.~A., Chuang, K.~J., Jäger, C., Ueberschaar, N., \& Henning, T. 2022, Nat. Astron., 6, 381

\bibitem[{Krasnokutski {et~al.}(2024)Krasnokutski, Jäger, Henning, Geffroy, Remaury, \& Poinot}]{Krasnokutski2024}
Krasnokutski, S.~A., Jäger, C., Henning, T., {et~al.} 2024, Sci. Adv., 10, eadj7179

\bibitem[{Mat{\'e} {et~al.}(2023)Mat{\'e}, Tanarro, Tim{\'o}n, Cernicharo, \& Herrero}]{mate2023}
Mat{\'e}, B., Tanarro, I., Tim{\'o}n, V., Cernicharo, J., \& Herrero, V.~J. 2023, MNRAS, stad1761

\bibitem[{Molpeceres {et~al.}(2024)Molpeceres, Tsuge, Furuya, Watanabe, San~Andrés, Rivilla, Colzi, \& Aikawa}]{Molpeceres24}
Molpeceres, G., Tsuge, M., Furuya, K., {et~al.} 2024, J. Phys. Chem. A, 128, 3874

\bibitem[{Potapov {et~al.}(2019)Potapov, J{\"a}ger, \& Henning}]{potapov2019}
Potapov, A., J{\"a}ger, C., \& Henning, T. 2019, ApJ, 880, 12

\bibitem[{Rachid {et~al.}(2021)Rachid, Brunken, De~Boe, Fedoseev, Boogert, \& Linnartz}]{rachid2021}
Rachid, M., Brunken, N., De~Boe, D., {et~al.} 2021, A\&A, 653, A116

\bibitem[{R{\"a}s{\"a}nen {et~al.}(1983)R{\"a}s{\"a}nen, Aspiala, Homanen, \& Murto}]{rasanen1983}
R{\"a}s{\"a}nen, M., Aspiala, A., Homanen, L., \& Murto, J. 1983, J. Mol. Struct., 96, 81

\bibitem[{Rieke {et~al.}(2015)Rieke, Wright, B{\"o}ker, Bouwman, Colina, Glasse, Gordon, Greene, G{\"u}del, Henning, {et~al.}}]{rieke2015}
Rieke, G.~H., Wright, G., B{\"o}ker, T., {et~al.} 2015, PASP, 127, 584

\bibitem[{Rivilla {et~al.}(2021)Rivilla, Jim{\'e}nez-Serra, Mart{\'\i}n-Pintado, Briones, Rodr{\'\i}guez-Almeida, Rico-Villas, Tercero, Zeng, Colzi, de~Vicente, {et~al.}}]{rivilla2021}
Rivilla, V.~M., Jim{\'e}nez-Serra, I., Mart{\'\i}n-Pintado, J., {et~al.} 2021, PNAS, 118, e2101314118

\bibitem[{Rocha {et~al.}(2024)Rocha, van Dishoeck, Ressler, van Gelder, Slavicinska, Brunken, Linnartz, Ray, Beuther, o~Garatti, {et~al.}}]{rocha2024}
Rocha, W., van Dishoeck, E., Ressler, M., {et~al.} 2024, A\&A, 683, A124

\bibitem[{Silva {et~al.}(1999)Silva, Duarte, \& Fausto}]{silva1999}
Silva, C.~F., Duarte, M. L.~T., \& Fausto, R. 1999, J. Mol. Struct., 482, 591

\bibitem[{Sladkova {et~al.}(2014)Sladkova, Lisovskaya, Sosnovskaya, Edimecheva, \& Shdyro}]{sladkova2014}
Sladkova, A.~A., Lisovskaya, A.~G., Sosnovskaya, A.~A., Edimecheva, I.~P., \& Shdyro, O.~I. 2014, Radiat. Phys. Chem., 96, 229

\bibitem[{Suhasaria {et~al.}(2015)Suhasaria, Thrower, \& Zacharias}]{suhasaria2015}
Suhasaria, T., Thrower, J., \& Zacharias, H. 2015, MNRAS, 454, 3317

\bibitem[{Thrower {et~al.}(2013)Thrower, Friis, Skov, Nilsson, Andersen, Ferrighi, J{\o}rgensen, Baouche, Balog, Hammer, {et~al.}}]{thrower2013}
Thrower, J.~D., Friis, E.~E., Skov, A.~L., {et~al.} 2013, J. Phys. Chem. C, 117, 13520

\bibitem[{Tseng {et~al.}(2010)Tseng, Chen, Wang, Peng, \& Lin}]{tseng2010}
Tseng, C.-L., Chen, Y.-K., Wang, S.-H., Peng, Z.-W., \& Lin, J.-L. 2010, J. Phys. Chem. C, 114, 11835

\bibitem[{Villadsen {et~al.}(2022)Villadsen, Ligterink, \& Andersen}]{villadsen2022}
Villadsen, T., Ligterink, N.~F., \& Andersen, M. 2022, A\&A, 666, A45

\bibitem[{{Vinogradoff} {et~al.}(2013){Vinogradoff}, {Duvernay}, {Danger}, {Theul{\'e}}, {Borget}, \& {Chiavassa}}]{vinogradoff2013}
{Vinogradoff}, V., {Duvernay}, F., {Danger}, G., {et~al.} 2013, \aap, 549, A40

\bibitem[{Wells {et~al.}(2015)Wells, Pel, Glasse, Wright, Aitink-Kroes, Azzollini, Beard, Brandl, Gallie, Geers, {et~al.}}]{wells2015}
Wells, M., Pel, J.-W., Glasse, A., {et~al.} 2015, PASP, 127, 646

\bibitem[{Wright {et~al.}(2015)Wright, Wright, Goodson, Rieke, Aitink-Kroes, Amiaux, Aricha-Yanguas, Azzollini, Banks, Barrado-Navascues, {et~al.}}]{wright2015}
Wright, G., Wright, D., Goodson, G., {et~al.} 2015, PASP, 127, 595

\end{thebibliography}
\end{document}